\documentstyle[psfig]{camera}

\begin{document}

%
\title{DETECTABILITY OF SPACE-TIME FLUCTUATIONS IN ULTRA HIGH 
ENERGY COSMIC RAY EXPERIMENTS }

%
\author{Roberto Aloisio$^1$, Pasquale Blasi$^2$, Angelo Galante$^3$,
\\ Piera Luisa Ghia$^{4,5}$ \And Aurelio F. Grillo$^1$}

%
\organization{\begin{center} 1. INFN - Laboratori Nazionali del Gran Sasso, SS. 17bis,
Assergi (L'Aquila) - Italy \\
2. INAF - Osservatorio Astrofisico di Arcetri, Largo E. Fermi 5,
50125 Firenze - Italy
 \\ 3. Dipartimento di Fisica, Universit\`a di L'Aquila, Via Vetoio
67100 Coppito (L'Aquila) - Italy 
\\4. CNR - IFSI, Sezione di Torino, Corso Fiume 4, 
10133 Torino - Italy
\\5. INFN - Sezione di Torino, 
Via P. Giuria 1, 10125 Torino - Italy \end{center}}

\maketitle

%
\noindent
{\bf ABSTRACT.} 
It is generally expected that quantum gravity affects the 
structure of space-time by introducing stochastic fluctuations in the
geometry, and, ultimately, in the measurements of four-distances and
four-momenta.
 These 
fluctuations may induce observable consequences on the propagation of 
ultra high energy particles, mainly in the range of energies of interest 
for cosmic ray physics, 
 over large distances, leading to their 
detection or constraining the underlying quantum gravitational structure.
We argue that the detectable effects of fluctuations may extend 
to energies much lower than the threshold for 
proton photopion production (the so-called GZK cut-off), so
 that lower energy observations may provide 
strong constraints on the role of a fluctuating space-time structure.

\section*{INTRODUCTION}
Since the advent of Special and General Relativity we are acquainted to the 
concept that space-time is dynamically related to its content. It is however 
generally thought as a preexisting, background arena in which our 
Universe lives.
However, as already recognized 45 years ago \cite{wheel}, quantum 
gravitational (QG) effects are likely to profoundly modify this picture: at the
Planck scale, which is the scale at which QG becomes important, the geometry 
of the space-time ceases to be definite and fluctuations are expected both 
in its geometrical and topological structure. 
In fact the most ambitious QG programs  do not stipulate a 
preexisting space-time which should instead emerge as the long 
(compared to Planck) distance
limit of some, more fundamental, quantum structure.

Although this is known since long time, it has always been relegated to the 
realm of the unreacheable super-Planckian world, {\it{i.e.}} at energies 
exceeding $10^{29}$ eV. 

This situation has rapidly changed in the last years with the realization that 
Nature provides us with probes (either UHE cosmic rays or gammas from 
very far, variable sources) which can feel  in their travel the fundamental
structure of 
our Universe. In this contribution we will focus on UHE particles. In this
case it has been realized that the onset of the processes 
responsible for  absorption of particles from very far sources on universal 
background radiation fields (of which the best studied is the Cosmic
Microwave Background Radiation, CMBR) is extremely sensitive to even tiny 
modifications of Lorentzian empty-space propagation, and that the 
corresponding absorption thresholds can be modified in a way testable even 
from present day experiments.

In previous work \cite{abgg} we focused attention to the effect of explicit 
modifications of Relativistic Invariance, compatible with all low energy
experiments and yet capable to be verified/falsified. In the present 
contribution we discuss the possible effects of the intrinsic uncertainty 
on {\it every} measurement induced by quantum gravity and discuss
in some details their implications. The results, when properly taking into 
account all processes affecting propagation, are striking and contrary to
intuition. The threshold for absorption effects moves to {\it lower} energies
with respect to the expected one\footnote{
In the following we will always use units such that  $\hbar=c=1$ unless 
needed.}.  
    
\section*{EFFECTS OF GEOMETRY FLUCTUATIONS ON PROPAGATION OF UHE
PARTICLES}

It is long known that QG effects preclude an arbitrary precision in the 
measurement of distances (time). There are various ways of demonstrating 
this fact,
but perhaps the simplest and most intuitive uses the operative definition of
distance, {\it i.e.}: $distance$ $\propto$ {\it N. of wavelengths between 
points}. To measure a distance $D \leq L_P$ (Planck length)  \footnote{We 
remind that $L_P=({{\hbar G} \over c^3})^{1\over 2} \approx 10^{-23} 
\quad cm$ is the Schwarzschild radius of a mass $M_P \approx 10^{28} \quad
eV$.} we need wavelenghts $ \lambda < L_P$, {\it i.e.} frequencies 
$> M_P$ thus 
leading inevitably to the formation of a black-hole, where information
is lost; in this sense distances $\leq L_P$ cannot be defined and the 
{\it minimum} uncertainty on distances is $\delta l \propto L_P$.

This uncertainty can be be transferred into uncertainty on momenta
assuming that this is also the uncertainty on the de-Broglie wavelength of a 
particle:
\begin{equation}
 \delta p = \delta {1 \over \lambda} = {1 \over \lambda^2} \delta \lambda =
{L_P \over \lambda ^2} = {p^2 \over M_p}.
\end{equation}
and similarly for energies.

This is the starting point for our considerations \cite{abggg}: we 
assume that a) Energies
and momenta fluctuate independently (but, only for simplicity, maintaining
rotational invariance, i.e. the space components of momenta fluctuate in the 
same way) and b) the relation between energy and 3-momentum (dispersion 
relation) also fluctuates independently\footnote{This last assumption is 
essentially equivalent to considering only {\it non conformal} metric
fluctuations \cite{camacho}.}:

\begin{equation}
E=\bar E + \alpha {\bar E^2 \over M_P} \quad
p=\bar p + \beta {\bar p^2 \over M_P} \quad
E^2-p^2= m^2 + \gamma {\bar E^3 \over M_P}
\end{equation}
where $\bar E, \bar p$ are the values obtained averaging over a large
number of measurements, and $\alpha, \beta, \gamma$ are random variables
with zero mean and variance $\approx 1$; the form of their distribution is
not important and for definiteness we assume it gaussian. 
Equation 2 expresses the fact that {\it each} measurement of energy and 
momenta is undetermined up to an uncertainty that grows with (energy) momenta
so that an infinite precision cannot be attained. On the other hand each 
interaction (through conservation of four-momentum)
is a measurement, so that
a particle propagating in a medium will have in each interaction a slightly
different momentum. In particular, particle having average momenta below
a certain absorption threshold would have a finite, non zero probability
to fluctuate above threshold and be absorbed.  

We now compute the effect of such fluctuations on UHE Cosmic Rays, and for
definitness we consider the propagation of UHE protons in the relic 
$3^oK$ radiation. In this case the absorption process is 
$p \gamma_{3^oK} \to \pi N$ with a threshold at $\approx 5 \cdot 10^{19}$ eV.
We compute the threshold for this process by stipulating {\it independent}
fluctuations for each initial and final particle, since 
typical scales of interactions are much larger than Planck ones. Solving
the conservation equations supplemented by the dispersion relations one
then obtains a distribution of threshold values, instead than a definite one
as in the normal case. 

To better clarify this case, consider a simplified (but still physical)
example, and assume that only relevant fluctuations act on the dispersion
relation of the UHE proton: this position is likely to introduce a relatively
small error since fluctuations increase with energy and final particles have 
lower energy. The conservation-dispersion equations are then as in \cite{abgg}
with the solution for the threshold (neglecting pion mass):

\begin{equation}
\gamma {p_0^3 \over {m_p^2 M_P}} {p_{th}^3 \over p_0^3} + {p_{th} 
\over p_0}-1=0
\end{equation} 
where $p_{th}$ is the threshold (with fluctuations) and $p_0$ the normal,
Lorentz invariant one. Clearly when $\gamma=0$ the threshold is equal to the 
normal one; on the other hand the coefficient of the cubic term is very large
($O(10^{13})$ for the case we consider) so, as soon as $\gamma \le -10^{-13}$, 
which happens almost exactly in the $50 \% $ of the cases for any reasonable,
symmetric 
distribution with zero  mean and unit variance, there is no 
solution, {\it i.e.} the reaction 
is not allowed. On the other hand, for $\gamma >10^{-13}$, 
an approximate solution is given by 
\begin{equation}
p_{th} \approx m_p^2 M_P^{1\over 3} \gamma^{-{1\over 3}}\approx 10^{-15}
\gamma^{-{1\over 3}}\quad {\rm eV}.
\end{equation}

Therefore, when 
$\gamma>0$, the probability of having a given threshold will peak around 
$\gamma\approx 1 \quad i.e. \quad p_{th} \approx 2\times 10^{-15}$ eV; 
for instance for a gaussian distribution we obtain a distribution of 
probability as in Figure 1 in the $50 \%$ of the cases in which the threshold
exists.
\begin{figure}[htb]
\begin{center}
  \psfig{file=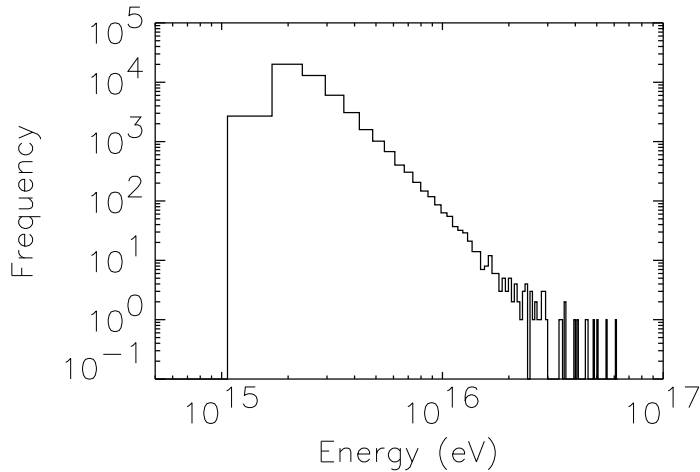,width=10.cm}
\caption{Distribution of probability of thresholds in the 50 \% of cases in
which the interaction is allowed.}
\end{center}
\end{figure}
In the case in which energies, momenta and dispersion relation of all particles
fluctuate independently the shape of the distribution is unchanged, only the 
probability that the interaction is kynematically forbidden decreases to
$\approx 30 \%$ since for fixed initial momenta the final ones can fluctuate
in a way to make the reaction allowed.

It is therefore clear that during propagation in the CMBR fluctuations 
of the space-time affect UHE particles in such a way that in a sizable 
fraction of cases the effective threshold is moved to lower energies
$\approx 2 \times 10^{15}$ eV. 

Before analyzing the experimental consequences of these distributions, it is 
interesting to derive the above effects in a more intuitive way, that can be 
most easily displayed in the case of pair production by UHE $\gamma$ on CMBR:
$\gamma(q) \gamma_{3^oK}(k) \to e^+e^-$. 
The four-momentum of  the initial photon will
fluctuate as in eq. 2, and we neglect the fluctuations on the low energy one. 
We further assume that once fluctuations in four-momenta are taken into 
account, then one can use normal Lorentz transformation to compute momenta in
different frames (\cite{lieu}). Then one can define, starting from the momenta
in Lab. frame for the above reaction,
a frame in which the space component of the total momenta  of the initial
particles is $k_i+q_i=0$, {\it i.e.} the ``rest'' frame. 
Clearly the velocity of
this frame with respect to the Laboratory will be affected by the fluctuations
of the energy and momentum:
\begin{equation}
\beta \approx 1-2 {k \over \bar q} +(\alpha - \beta){\bar q \over M_P}
\end{equation}
and this parameter becomes unphysical when $\beta >1$; given the 
form of fluctuations, this happens in almost exactly the $50 \%$ of the
cases as above.

Using this parameter we can now compute the total energy of the initial 
state in the ``rest'' frame; the reaction then is allowed if this 
energy is larger than 
the sum of the masses of the final particles\footnote{If also
momenta of final particles are allowed to fluctuate in fact this condition 
could be somewhat weakened, leading to a larger probability of interaction.}: 
\begin{equation}
E_{CM} \approx \left (k \bar q - (\alpha-\beta) {\bar q^3 \over {2 M_P}}
\right )^{1 \over 2} \ge 2 m_e
\end{equation}
from which 
\begin{equation}
\bar q \ge \bar q_{thr}\approx \left ( {2 m^2_e M_P \over {\alpha - \beta}} 
\right )^{1 \over 3} \approx 10^{-12} (\alpha-\beta)^{-{1\over 3}} {\rm eV}
\end{equation}
which has the same form of the analogous threshold (eq. 4) in the case of 
UHE proton propagation, displaying in a clear and intuitive way the origin
of the fluctuations for the absorption thresholds.

\section*{EXPERIMENTAL CONSEQUENCES}

To analize the detectable consequences of the above findings we need to
consider the propagation of UHE particles (protons). In {\it each} interaction,
for energies $>2 \cdot 10^{15}$ eV,
the probability of being absorbed is $P_{int} \approx 70 \%$ and of escaping
$P_{esc}=1-P_{int}$. We further assume that the cross section (and interaction
length $\lambda$) for absorption
is unchanged with respect to the L.I. case (although in a different region of
energy) at least far above the new threshold. Therefore the
effective interaction length is slightly increased $\lambda_{eff}
\approx {\lambda \over P_{int}}$, and for propagation
over distances $L >> \lambda$ we have $P^T_{esc}=P_{esc}^{L\over \lambda} \to
0$. Therefore, although in a single interaction $\approx 30 \%$ of particles
escape absorption, in the long run essentially all are absorbed although with
a sligthtly increased interaction length.

The situation is essentially the same as for the GZK cutoff in the L.I. 
approach, the striking difference being that the GZK feature is now moved
to much lower energies, and there is no more 
anything special with $10^{20}$ eV.

Is this prediction testable? Clearly it is easily falsifiable: if the GZK 
feature would be detected, than the above is wrong, and this implies that some
of the assumptions are untenable, for instance the fluctuations might be
'conformal' with much smaller effects, or their variance $< 10^{-13}$ or again
proportional to $(p/M_P)^\alpha$ with $\alpha>2.3$. 
In all these cases this would give important hints 
for QG model building. 

On the other hand if future experiments will not 
find the GZK feature where normally expected, then  our 
prediction is that the {\it entire} extragalactic component of CRs above 
$\approx 2 \cdot 10^{15}$ eV, not limited to those exceeding $10^{20}$ eV,  
arrives at our detectors from within a (slightly enlarged) GZK horizon.
This prediction might have observable effects, for instance in terms of 
anisotropy of UHECR sources; however a detailed propagation study is needed.
On the other hand particles that suffered interaction would pile up
around $10^{15}$ eV, but for protons their detectability is questionable, 
due to the much larger abundance of galactic CRs. However photons produced
in the decay of secondary neutral pions would initiate a cascade on CMBR
and pile up at GeV energies, with a predicted flux orders of magnitude larger 
than the EGRET experimental limit, and already excluded; but this conclusion 
relies on L.I. decay lengths and cascade developement estimates, 
which could be modified by
fluctuations, so again a detailed propagation analysis has to be performed.
In relation to this, it is interesting to notice that the use of standard
Lorentz transformations on fluctuating momenta could in principle allow to 
take into account space-time fluctuations within standard propagation
codes.

\end{document}